\documentclass[conference]{IEEEtran}
\usepackage[nopdftrailerid=true]{pdfprivacy}
\usepackage{multirow}
\usepackage{siunitx}
\usepackage{diagbox}
\usepackage{amssymb}
\usepackage{hyperref}
\usepackage{subcaption}
\usepackage{adjustbox}
\usepackage{tabularx}
\usepackage{booktabs}
\usepackage{makecell}
\usepackage{tikz}

\IEEEoverridecommandlockouts{}

\begin{document}

\title{Signal2Image Modules in Deep Neural Networks for EEG Classification}
\author{Paschalis Bizopoulos, George I. Lambrou and Dimitrios Koutsouris%
\thanks{The authors are with Biomedical Engineering Laboratory, School of Electrical and Computer Engineering, National Technical University of Athens, Athens 15780, Greece e-mail: {\tt\small pbizop@gmail.com}, {\tt\small glamprou@med.uoa.gr}, {\tt\small dkoutsou@biomed.ntua.gr.}}
}

\maketitle

\begin{abstract}
	Deep learning has revolutionized computer vision utilizing the increased availability of big data and the power of parallel computational units such as graphical processing units.
	The vast majority of deep learning research is conducted using images as training data, however the biomedical domain is rich in physiological signals that are used for diagnosis and prediction problems.
	It is still an open research question how to best utilize signals to train deep neural networks.

	In this paper we define the term Signal2Image (S2Is) as trainable or non-trainable prefix modules that convert signals, such as Electroencephalography (EEG), to image-like representations making them suitable for training image-based deep neural networks defined as `base models'.
	We compare the accuracy and time performance of four S2Is (`signal as image', spectrogram, one and two layer Convolutional Neural Networks (CNNs)) combined with a set of `base models' (LeNet, AlexNet, VGGnet, ResNet, DenseNet) along with the depth-wise and 1D variations of the latter.
	We also provide empirical evidence that the one layer CNN S2I performs better in eleven out of fifteen tested models than non-trainable S2Is for classifying EEG signals and present visual comparisons of the outputs of some of the S2Is.
\end{abstract}

\section{INTRODUCTION}
Most methods for solving biomedical problems until recently involved handcrafting features and trying to mimic human experts, which is increasingly proven to be inefficient and error-prone.
Deep learning is emerging as a powerful solution for a wide range of problems in biomedicine achieving superior results compared to traditional machine learning.
The main advantage of methods that use deep learning is that they automatically learn hierarchical features from training data making them scalable and generalizable.
This is achieved with the use of multilayer networks, that consist of million parameters~\cite{krizhevsky2012imagenet}, trained with backpropagation~\cite{rumelhart1986learning} on large amount of data.
Although deep learning is mainly used in biomedical images there is also a wide range of physiological signals, such as Electroencephalography (EEG), that are used for diagnosis and prediction problems.
EEG is the measure of the electrical field produced by the brain and is used for sleep pattern classification~\cite{aboalayon2016sleep}, brain computer interfaces~\cite{al2017review}, cognitive/affective monitoring~\cite{lotte1999electroencephalography} and epilepsy identification~\cite{acharya2013automated}.

Yannick et al.~\cite{yannick2019deep} reviewed deep learning studies using EEG and have identified a general increase in accuracy when deep learning and specifically Convolutional Neural Networks (CNNs) are used instead of traditional machine learning methods.
However, they do not mention which specific characteristics of CNN architectures are indicated to increase performance.
It is still an open research question how to best use EEG for training deep learning models.

One common approach that previous studies have used for classifying EEG signals was feature extraction from the frequency and time-frequency domains utilizing the theory behind EEG band frequencies~\cite{langkvist2012sleep}: delta (0.5--4 Hz), theta (4--8 Hz), alpha (8--13 Hz), beta (13--20 Hz) and gamma (20--64 Hz).
Truong et al.~\cite{truong2018convolutional} used Short-Time Fourier Transform (STFT) on a 30 second sliding window to train a three layer CNN on stacked time-frequency representations for seizure prediction and evaluated their method on three EEG databases.
Khan et al.~\cite{khan2018focal} transformed the EEGs in time-frequency domain using multi-scale wavelets and then trained a six layer CNN on these stacked multi-scale representations for predicting the focal onset seizure demonstrating promising results.

Feature extraction from the time-frequency domain has also been used for other EEG related tasks, besides epileptic seizure prediction.
Zhang et al.~\cite{zhang2017pattern} trained an ensemble of CNNs containing two to ten layers using STFT features extracted from EEG band frequencies for mental workload classification.
Giri et al.~\cite{giri2016ischemic} extracted statistical and information measures from frequency domain to train an 1D CNN with two layers to identify ischemic stroke.

For the purposes of this paper and for easier future reference we define the term Signal2Image module (S2I) as any module placed after the raw signal input and before a `base model' which is usually an established architecture for imaging problems.
An important property of a S2I is whether it consists of trainable parameters such as convolutional and linear layers or it is non-trainable such as traditional time-frequency methods.
Using this definition we can also derive that most previous methods for EEG classification use non-trainable S2Is and that no previous study has compared trainable with non-trainable S2Is.

In this paper we compare non-trainable and trainable S2Is combined with well known `base models' neural network architectures along with the 1D and depth-wise variations of the latter.
A high level overview of these combined methods is shown in Fig.~\ref{fig:highleveloverview}.
Although we choose the EEG epileptic seizure recognition dataset from University of California, Irvine (UCI)~\cite{andrzejak2001indications} for EEG classification, the implications of this study could be generalized in any kind of signal classification problem.
Here we also refer to CNN as a neural network consisting of alternating convolutional layers each one followed by a Rectified Linear Unit (ReLU) and a max pooling layer and a fully connected layer at the end while the term `layer' denotes the number of convolutional layers.

\section{DATA}
\begin{figure}[!t]
	\centering
	\begin{tikzpicture}[]
		\node[] at (-4.5, 0){\includegraphics[scale=0.2,angle=90]{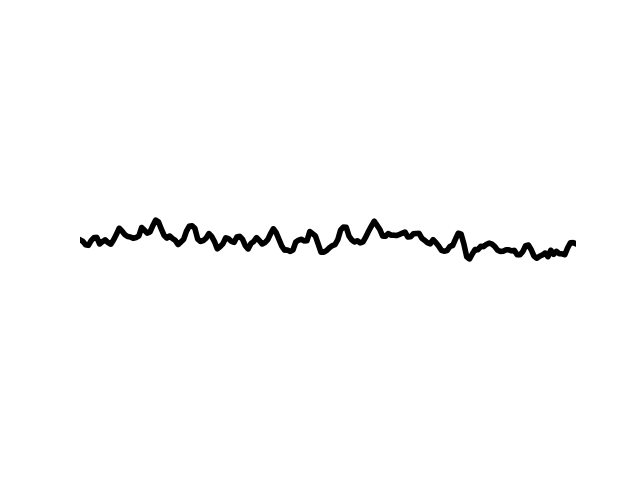}};
		\node[align=left] at (-4.2, 0) {$x_i$};
		\draw[dashed,->] (-4, 0) -- (-3.8, 0);
		\node[draw, minimum height=2.55cm] at (-3.5, 0) {$m$};
		\draw[dashed,->] (-3.2, 0) -- (-3, 0);
		\node[] at (-1.7, 0){\includegraphics[scale=0.41,angle=90]{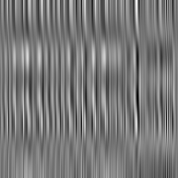}};
		\draw[dashed,->] (-0.4, 0) -- (-0.2, 0);
		\node[draw, minimum height=2.55cm] at (0.1, 0) {$b_{d}$};
		\draw[dashed,->] (0.4, 0) -- (0.6, 0);
		\node[align=left] at (0.8, 0) {$\hat{y_i}$};
		\node[minimum width=0.5cm, minimum height=0.5cm] at (1.67, 1.1) {\footnotesize Open 0.1\%};
		\node[minimum width=0.5cm, minimum height=0.5cm] at (1.74, 0.55) {\footnotesize Closed 0.2\%};
		\node[minimum width=0.5cm, minimum height=0.5cm] at (1.78, 0) {\footnotesize Healthy 0.9\%};
		\node[minimum width=0.5cm, minimum height=0.5cm] at (1.8, -0.55) {\footnotesize Tumor 34.7\%};
		\node[minimum width=0.5cm, minimum height=0.5cm] at (1.9, -1.1) {\footnotesize Epilepsy 64.1\%};
	\end{tikzpicture}
	\caption{High level overview of a feed-forward pass of the combined methods.
	$x_i$ is the input, $m$ is the Signal2Image module, $b_{d}$ is the 1D or 2D architecture `base model' for $d=1,2$ respectively and $\hat{y_i}$ is the predicted output.
	The names of the classes are depicted at the right along with the predictions for this example signal.
	The image between $m$ and $b_{d}$ depicts the output of the one layer CNN Signal2Image module, while the `signal as image' and spectrogram have intermediate images as those depicted at the second and third row of Fig.~\ref{fig:signal2imageoutputs}.
	Arrows denote the flow of the feed-forward pass.
	For the 1D architectures $m$ is omitted and no intermediate image is generated.}\label{fig:highleveloverview}
\end{figure}
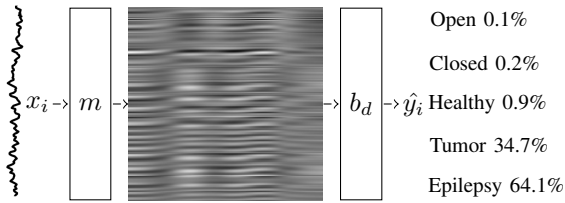

The UCI EEG epileptic seizure recognition dataset~\cite{andrzejak2001indications} consists of $500$ signals each one with $4097$ samples (23.5 seconds).
The dataset is annotated into five classes with $100$ signals for each class (in parenthesis the shortened class names used in Fig.~\ref{fig:highleveloverview} and~\ref{fig:signal2imageoutputs}):
\begin{enumerate}
	\item healthy patient while having his eyes open (Open),
	\item healthy patient while having his eyes closed (Closed),
	\item patient with tumor taken from healthy area (Healthy),
	\item patient with tumor taken from tumor area (Tumor),
	\item patient while having seizure activity (Epilepsy)
\end{enumerate}

For the purposes of this paper we use a variation of the database\footnote{\url{https://archive.ics.uci.edu/ml/datasets/Epileptic+Seizure+Recognition}} in which the EEG signals are split into segments with $178$ samples each, resulting in a balanced dataset that consists of $11500$ EEG signals.

\section{METHODS}
\subsection{Definitions}
We define the dataset $D=\{x_i, y_i\}_{i=1\ldots N}$ where $x_i \in \mathbb{Z}^n$ and $y_i \in \{1, 2, 3, 4, 5\}$ denote the $i^{th}$ input signal with dimensionality $n=178$ and the $i^{th}$ class with five possible classes respectively. 
$N=11500$ is the number of observations.

We also define the set of S2Is as $M$ and the member of this set as $m$ which include the following modules:
\begin{itemize}
	\item `signal as image' (non-trainable)
	\item spectrogram (non-trainable)
	\item one and two layers CNN (trainable)
\end{itemize}

We then define the set of `base models' as $B$ and the member of this set as $b_d$ where $d=[1,2]$ denotes the dimensionality of the convolutional, max pooling, batch normalization and adaptive average pooling layers.
$B$ includes the following $b_d$ along with their depth-wise variations and their equivalent 1D architectures for $d=1$ (for a complete list refer to first two rows of Table.~\ref{table:results}):
\begin{itemize}
	\item LeNet~\cite{lecun1998gradient}
	\item AlexNet~\cite{krizhevsky2012imagenet}
	\item VGGnet~\cite{simonyan2014very}
	\item ResNet~\cite{he2016deep}
	\item DenseNet~\cite{huang2017densely}
\end{itemize}

We finally define the combinations of $m$ and $b_d$ as the members $c$ of the set of combined models $C$.
Using the previous definitions, the aim of this paper is the evaluation of the set of models $C$, where $C$ is the combined set of $M$ and $B$ i.e. $C=M\times B$ w.r.t.\ time performance and class accuracy trained on $D$.

\subsection{Signal2Image Modules}
In this section we describe the internals of each S2I module.
For the `signal as image' module, we normalized the amplitude of $x_i$ to the range $[1, 178]$.
The results were inverted along the y-axis, rounded to the nearest integer and then they were used as the y-indices for the pixels with amplitude $255$ on a $178\times 178$ image initialized with zeros.

For the spectrogram module, which is used for visualizing the change of the frequency of a non-stationary signal over time~\cite{oppenheim1999discrete}, we used a Tukey window with a shape parameter of $0.25$, a segment length of $8$ samples, an overlap between segments of $4$ samples and a fast Fourier transform of $64$ samples to convert the $x_i$ into the time-frequency domain.
The resulted spectrogram, which represents the magnitude of the power spectral density ($V^2/Hz$) of $x_i$, was then upsampled to $178\times 178$ using bilinear pixel interpolation.

For the CNN modules with one and two layers, $x_i$ is converted to an image using learnable parameters instead of some static procedure.
The one layer module consists of one 1D convolutional layer (kernel sizes of $3$ with $8$ channels).
The two layer module consists of two 1D convolutional layers (kernel sizes of $3$ with $8$ and $16$ channels) with the first layer followed by a ReLU activation function and a 1D max pooling operation (kernel size of $2$).
The feature maps of the last convolutional layer for both modules are then concatenated along the y-axis and then resized to $178\times 178$ using bilinear interpolation.

\begin{figure*}[!t]
	\centering
	\rotatebox[origin=l]{90}{\small Original Signal}
	\begin{subfigure}{0.19\linewidth}
		\centering
		\includegraphics[scale=0.2]{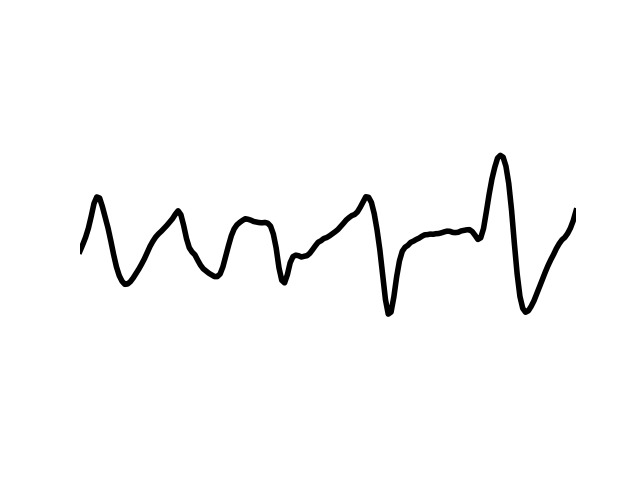}
	\end{subfigure}
	\begin{subfigure}{0.19\linewidth}
		\centering
		\includegraphics[scale=0.2]{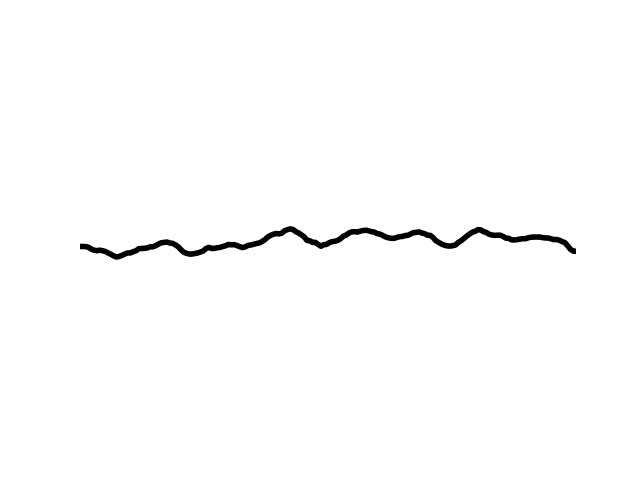}
	\end{subfigure}
	\begin{subfigure}{0.19\linewidth}
		\centering
		\includegraphics[scale=0.2]{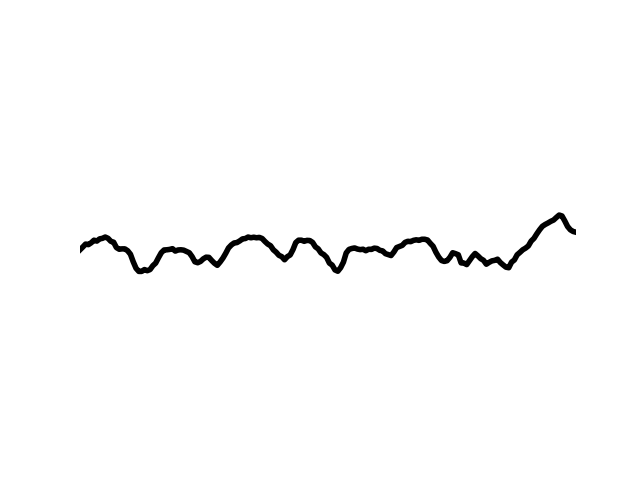}
	\end{subfigure}
	\begin{subfigure}{0.19\linewidth}
		\centering
		\includegraphics[scale=0.2]{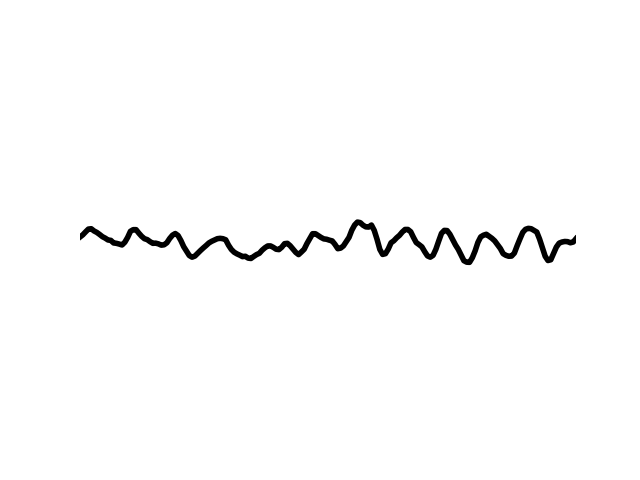}
	\end{subfigure}
	\begin{subfigure}{0.19\linewidth}
		\centering
		\includegraphics[scale=0.2]{python/tmp/signal-epilepsy.png}
	\end{subfigure}

	\rotatebox[origin=l]{90}{\small `Signal as Image'}
	\begin{subfigure}{0.19\linewidth}
		\centering
		\includegraphics[scale=0.5]{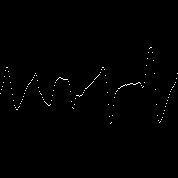}
	\end{subfigure}
	\begin{subfigure}{0.19\linewidth}
		\centering
		\includegraphics[scale=0.5]{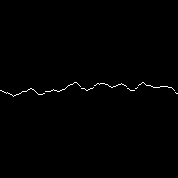}
	\end{subfigure}
	\begin{subfigure}{0.19\linewidth}
		\centering
		\includegraphics[scale=0.5]{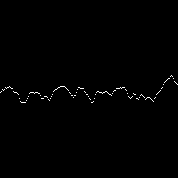}
	\end{subfigure}
	\begin{subfigure}{0.19\linewidth}
		\centering
		\includegraphics[scale=0.5]{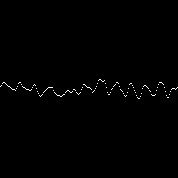}
	\end{subfigure}
	\begin{subfigure}{0.19\linewidth}
		\centering
		\includegraphics[scale=0.5]{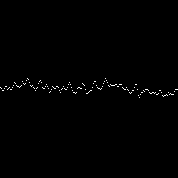}
	\end{subfigure}

	\rotatebox[origin=l]{90}{\small \hspace{1mm} Spectrogram}
	\begin{subfigure}{0.19\linewidth}
		\centering
		\includegraphics[scale=0.5]{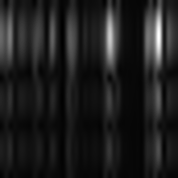}
	\end{subfigure}
	\begin{subfigure}{0.19\linewidth}
		\centering
		\includegraphics[scale=0.5]{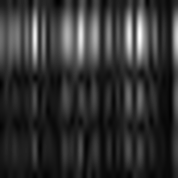}
	\end{subfigure}
	\begin{subfigure}{0.19\linewidth}
		\centering
		\includegraphics[scale=0.5]{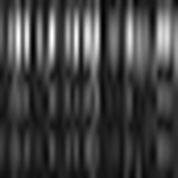}
	\end{subfigure}
	\begin{subfigure}{0.19\linewidth}
		\centering
		\includegraphics[scale=0.5]{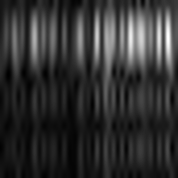}
	\end{subfigure}
	\begin{subfigure}{0.19\linewidth}
		\centering
		\includegraphics[scale=0.5]{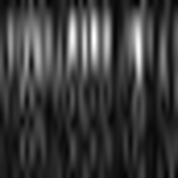}
	\end{subfigure}

	\rotatebox[origin=l]{90}{\small \hspace{5mm} one layer CNN}
	\begin{subfigure}{0.19\linewidth}
		\centering
		\includegraphics[scale=0.5]{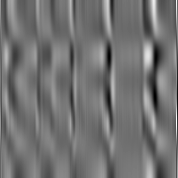}
		\caption{Open}
	\end{subfigure}
	\begin{subfigure}{0.19\linewidth}
		\centering
		\includegraphics[scale=0.5]{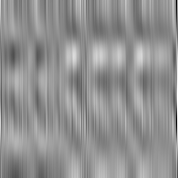}
		\caption{Closed}
	\end{subfigure}
	\begin{subfigure}{0.19\linewidth}
		\centering
		\includegraphics[scale=0.5]{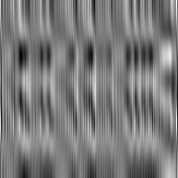}
		\caption{Healthy}
	\end{subfigure}
	\begin{subfigure}{0.19\linewidth}
		\centering
		\includegraphics[scale=0.5]{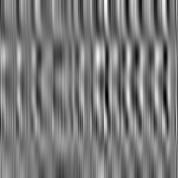}
		\caption{Tumor}
	\end{subfigure}
	\begin{subfigure}{0.19\linewidth}
		\centering
		\includegraphics[scale=0.5]{python/tmp/cnn-epilepsy.png}
		\caption{Epilepsy}
	\end{subfigure}
	\caption{Visualizations of the original signals and the outputs of the S2Is for each class.
	The x, y-axis of the first row are in \SI{}{\micro{} V} and time samples respectively.
	The x, y-axis of the rest of the subfigures denote spatial information, since we do not inform the `base model' the concept of time along the x-axis or the concept of frequency along the y-axis.
	Higher pixel intensity denotes higher amplitude.}\label{fig:signal2imageoutputs}
\end{figure*}

We constrain the output for all $m$ to a $178\times 178$ image to enable visual comparison.
Three identical channels were also stacked for all $m$ outputs to satisfy the input size requirements for $b_d$.
Architectures of all $b_d$ remained the same, except for the number of the output nodes of the last linear layer which was set to five to correspond to the number of classes of $D$.
An example of the respective outputs of some of the $m$ (the one/two layers CNN produced similar visualizations) are depicted in the second, third and fourth row of Fig.~\ref{fig:signal2imageoutputs}.

\subsection{Evaluation}
Convolutional layers of $m$ were initialized using Kaiming uniform~\cite{he2015delving}. Values are sampled from the uniform distribution $\mathcal{U}(-c, c)$, where $c$ is:
\begin{equation}
	c = \sqrt{\frac{6}{(1 + a^2) k}}
\end{equation}

\noindent
,$a$ in this study is set to zero and $k$ is the size of the input of the layer.
The linear layers of $m$ were initialized using $\mathcal{U}(-\frac{1}{\sqrt{k}},\frac{1}{\sqrt{k}})$.
The convolutional and linear layers of all $b_d$ were initialized according to their original implementation.

We used Adam~\cite{kingma2014adam} as the optimizer with learning rate $lr=0.001$, betas $b_1=0.9$, $b_2=0.999$, epsilon $\epsilon=10^{-8}$ without weight decay and cross entropy as the loss function.
Batch size was $20$ and no additional regularization was used besides the structural such as dropout layers that some of the `base models' (AlexNet, VGGnet and DenseNet) have.

Out of the 11500 signals we used $76\%$, $12\%$ and $12\%$ of the data ($8740,1380,1380$ signals) as training, validation and test data respectively.
No artifact handling or preprocessing was performed.
All networks were trained for $100$ epochs and model selection was performed using the best validation accuracy out of all the epochs.
We used PyTorch~\cite{paszke2017automatic} for implementing the neural network architectures and training/preprocessing was done using a NVIDIA Titan X Pascal GPU 12GB RAM and a 12 Core Intel i7--8700 CPU @ 3.20GHz on a Linux operating system.

\begin{table*}[!t]
	\caption{Test accuracies (\%) for combined models.
	The second row indicates the number of layers.
	Bold indicates the best accuracy for each base model.}\label{table:results}
	\begin{adjustbox}{width=\textwidth}
		\begin{tabular}{lccccccccccccccc}
\toprule
{} &  lenet &  alexnet &  vgg11 &  vgg13 &  vgg16 &  vgg19 &  resnet18 &  resnet34 &  resnet50 &  resnet101 &  resnet152 &  densenet121 &  densenet161 &  densenet169 &  densenet201 \\
\midrule
\textbf{1D, }                & 72.6          & 78.8          & 76.9                        & \textbf{79.0}               & 79.5                         & \textbf{79.3} & 81.5          & 82.5          & 81.4          & 78.8          & 81.4          & 81.8          & \textbf{83.3} & 82.1          & 82.0          \\
\textbf{2D, signal as image} & 67.9          & 68.3          & 74.1                        & 74.7                        & 72.7                         & 72.5          & 73.3          & 71.7          & 74.1          & 72.3          & 74.1          & 74.7          & 72.5          & 75.2          & 75.0          \\
\textbf{2D, spectrogram    } & 73.2          & 74.0          & 77.9                        & 76.3                        & 77.5                         & 76.0          & 76.2          & 79.0          & 77.2          & 74.6          & 75.3          & 74.1          & 75.2          & 77.0          & 75.4          \\
\textbf{2D, one layer CNN  } & \textbf{75.8} & \textbf{82.0} & \textbf{84.0}               & 77.9                        & 80.7                         & 78.4          & \textbf{85.1} & \textbf{84.6} & \textbf{83.0} & \textbf{85.0} & \textbf{83.3} & \textbf{84.3} & 80.7          & \textbf{85.0} & \textbf{85.3} \\
\textbf{2D, two layer CNN  } & 75.0          & 77.9          & 80.7                        & 78.8                        & \textbf{81.1}                & 74.9          & 78.3          & 80.0          & 78.3          & 77.1          & 80.9          & 83.2          & 82.3          & 79.0          & 79.1          \\
\bottomrule
\end{tabular}

	\end{adjustbox}
\end{table*}

\section{RESULTS}
As shown in Table.~\ref{table:results} the one layer CNN DenseNet201 achieved the best accuracy of $85.3\%$ with training time 70 seconds/epoch on average.
In overall the one layer CNN S2I achieved best accuracies for eleven out of fifteen `base models'.
The two layer CNN S2I achieved worse even compared with the 1D variants, indicating that increase of the S2I depth is not beneficial.
The `signal as image' and spectrogram S2Is performed much worse than 1D variants and the CNN S2Is.
The spectrogram S2I results are in contrary with the expectation that the interpretable time-frequency representation would help in finding good features for classification.
We hypothesize that the spectrogram S2I was hindered by its lack of non-trainable parameters.
Another outcome of these experiments is that increasing the depth of the base models did not increase the accuracy which is inline with previous results~\cite{schirrmeister2017deep}.

\section{CONCLUSIONS}
In this paper we have shown empirical evidence that 1D `base model' variations and trainable S2Is (especially the one layer CNN) perform better than non-trainable S2Is.
However more work needs to be done for full replacing non-trainable S2Is, not only from the scope of achieving higher accuracy results but also increasing the interpretability of the model.
Another point of reference is that the combined models were trained from scratch based on the hypothesis that pretrained low level features of the `base models' might not be suitable for spectrogram-like images such as those created by S2Is.
Future work could include testing this hypothesis by initializing a `base model' using transfer learning or other initialization methods.
Moreover, trainable S2Is and 1D `base model' variations could also be used for other physiological signals besides EEG such as Electrocardiography, Electromyography and Galvanic Skin Response.

\section*{ACKNOWLEDGMENT}
This work was supported by the European Union's Horizon 2020 research and innovation programme under Grant agreement 769574.
We gratefully acknowledge the support of NVIDIA with the donation of the Titan X Pascal GPU used for this research.

\bibliographystyle{IEEEtran}
\bibliography{ms.bib}

\begin{thebibliography}{10}
\providecommand{\url}[1]{#1}
\csname url@samestyle\endcsname
\providecommand{\newblock}{\relax}
\providecommand{\bibinfo}[2]{#2}
\providecommand{\BIBentrySTDinterwordspacing}{\spaceskip=0pt\relax}
\providecommand{\BIBentryALTinterwordstretchfactor}{4}
\providecommand{\BIBentryALTinterwordspacing}{\spaceskip=\fontdimen2\font plus
\BIBentryALTinterwordstretchfactor\fontdimen3\font minus
  \fontdimen4\font\relax}
\providecommand{\BIBforeignlanguage}[2]{{%
\expandafter\ifx\csname l@#1\endcsname\relax
\typeout{** WARNING: IEEEtran.bst: No hyphenation pattern has been}%
\typeout{** loaded for the language `#1'. Using the pattern for}%
\typeout{** the default language instead.}%
\else
\language=\csname l@#1\endcsname
\fi
#2}}
\providecommand{\BIBdecl}{\relax}
\BIBdecl

\bibitem{krizhevsky2012imagenet}
A.~Krizhevsky, I.~Sutskever, and G.~E. Hinton, ``Imagenet classification with
  deep convolutional neural networks,'' in \emph{Advances in neural information
  processing systems}, 2012, pp. 1097--1105.

\bibitem{rumelhart1986learning}
D.~E. Rumelhart, G.~E. Hinton, and R.~J. Williams, ``Learning representations
  by back-propagating errors,'' \emph{nature}, vol. 323, no. 6088, p. 533,
  1986.

\bibitem{aboalayon2016sleep}
K.~Aboalayon, M.~Faezipour, W.~Almuhammadi, and S.~Moslehpour, ``Sleep stage
  classification using eeg signal analysis: a comprehensive survey and new
  investigation,'' \emph{Entropy}, vol.~18, no.~9, p. 272, 2016.

\bibitem{al2017review}
A.~Al-Nafjan, M.~Hosny, Y.~Al-Ohali, and A.~Al-Wabil, ``Review and
  classification of emotion recognition based on eeg brain-computer interface
  system research: a systematic review,'' \emph{Applied Sciences}, vol.~7,
  no.~12, p. 1239, 2017.

\bibitem{lotte1999electroencephalography}
F.~Lotte, L.~Bougrain, and M.~Clerc, ``Electroencephalography (eeg)-based
  brain--computer interfaces,'' \emph{Wiley Encyclopedia of Electrical and
  Electronics Engineering}, pp. 1--20, 1999.

\bibitem{acharya2013automated}
U.~R. Acharya, S.~V. Sree, G.~Swapna, R.~J. Martis, and J.~S. Suri, ``Automated
  eeg analysis of epilepsy: a review,'' \emph{Knowledge-Based Systems},
  vol.~45, pp. 147--165, 2013.

\bibitem{yannick2019deep}
R.~Yannick, B.~Hubert, A.~Isabela, G.~Alexandre, F.~Jocelyn \emph{et~al.},
  ``Deep learning-based electroencephalography analysis: a systematic review,''
  \emph{arXiv preprint arXiv:1901.05498}, 2019.

\bibitem{langkvist2012sleep}
M.~L{\"a}ngkvist, L.~Karlsson, and A.~Loutfi, ``Sleep stage classification
  using unsupervised feature learning,'' \emph{Advances in Artificial Neural
  Systems}, vol. 2012, p.~5, 2012.

\bibitem{truong2018convolutional}
N.~D. Truong, A.~D. Nguyen, L.~Kuhlmann, M.~R. Bonyadi, J.~Yang, S.~Ippolito,
  and O.~Kavehei, ``Convolutional neural networks for seizure prediction using
  intracranial and scalp electroencephalogram,'' \emph{Neural Networks}, vol.
  105, pp. 104--111, 2018.

\bibitem{khan2018focal}
H.~Khan, L.~Marcuse, M.~Fields, K.~Swann, and B.~Yener, ``Focal onset seizure
  prediction using convolutional networks,'' \emph{IEEE Transactions on
  Biomedical Engineering}, vol.~65, no.~9, pp. 2109--2118, 2018.

\bibitem{zhang2017pattern}
J.~Zhang, S.~Li, and R.~Wang, ``Pattern recognition of momentary mental
  workload based on multi-channel electrophysiological data and ensemble
  convolutional neural networks,'' \emph{Frontiers in neuroscience}, vol.~11,
  p. 310, 2017.

\bibitem{giri2016ischemic}
E.~P. Giri, M.~I. Fanany, A.~M. Arymurthy, and S.~K. Wijaya, ``Ischemic stroke
  identification based on eeg and eog using id convolutional neural network and
  batch normalization,'' in \emph{Advanced Computer Science and Information
  Systems (ICACSIS), 2016 International Conference on}.\hskip 1em plus 0.5em
  minus 0.4em\relax IEEE, 2016, pp. 484--491.

\bibitem{andrzejak2001indications}
R.~G. Andrzejak, K.~Lehnertz, F.~Mormann, C.~Rieke, P.~David, and C.~E. Elger,
  ``Indications of nonlinear deterministic and finite-dimensional structures in
  time series of brain electrical activity: Dependence on recording region and
  brain state,'' \emph{Physical Review E}, vol.~64, no.~6, p. 061907, 2001.

\bibitem{lecun1998gradient}
Y.~LeCun, L.~Bottou, Y.~Bengio, and P.~Haffner, ``Gradient-based learning
  applied to document recognition,'' \emph{Proceedings of the IEEE}, vol.~86,
  no.~11, pp. 2278--2324, 1998.

\bibitem{simonyan2014very}
K.~Simonyan and A.~Zisserman, ``Very deep convolutional networks for
  large-scale image recognition,'' \emph{arXiv preprint arXiv:1409.1556}, 2014.

\bibitem{he2016deep}
K.~He, X.~Zhang, S.~Ren, and J.~Sun, ``Deep residual learning for image
  recognition,'' in \emph{Proceedings of the IEEE conference on computer vision
  and pattern recognition}, 2016, pp. 770--778.

\bibitem{huang2017densely}
G.~Huang, Z.~Liu, L.~Van Der~Maaten, and K.~Q. Weinberger, ``Densely connected
  convolutional networks,'' in \emph{Proceedings of the IEEE conference on
  computer vision and pattern recognition}, 2017, pp. 4700--4708.

\bibitem{oppenheim1999discrete}
A.~V. Oppenheim, \emph{Discrete-time signal processing}.\hskip 1em plus 0.5em
  minus 0.4em\relax Pearson Education India, 1999.

\bibitem{he2015delving}
K.~He, X.~Zhang, S.~Ren, and J.~Sun, ``Delving deep into rectifiers: Surpassing
  human-level performance on imagenet classification,'' in \emph{Proceedings of
  the IEEE international conference on computer vision}, 2015, pp. 1026--1034.

\bibitem{kingma2014adam}
D.~P. Kingma and J.~Ba, ``Adam: A method for stochastic optimization,''
  \emph{arXiv preprint arXiv:1412.6980}, 2014.

\bibitem{paszke2017automatic}
A.~Paszke, S.~Gross, S.~Chintala, G.~Chanan, E.~Yang, Z.~DeVito, Z.~Lin,
  A.~Desmaison, L.~Antiga, and A.~Lerer, ``Automatic differentiation in
  pytorch,'' 2017.

\bibitem{schirrmeister2017deep}
R.~T. Schirrmeister, J.~T. Springenberg, L.~D.~J. Fiederer, M.~Glasstetter,
  K.~Eggensperger, M.~Tangermann, F.~Hutter, W.~Burgard, and T.~Ball, ``Deep
  learning with convolutional neural networks for eeg decoding and
  visualization,'' \emph{Human brain mapping}, vol.~38, no.~11, pp. 5391--5420,
  2017.

\end{thebibliography}

\end{document}